\documentclass[10pt,english,superscriptaddress,epsfig,nofootinbib,preprintnumbers]{revtex4}
\usepackage[T1]{fontenc}
\usepackage[latin1]{inputenc}
\usepackage{graphicx}

\makeatletter

\makeatletter

\usepackage{dcolumn}

\usepackage{bm}

\usepackage{longtable}

\usepackage{epstopdf}

\makeatletter



\makeatother

\graphicspath{{./images/}}

\makeatother

\usepackage{babel}
\makeatother
\begin{document}
\bibliographystyle{apsrev}

\title{Peculiar Acceleration}

\author{Luca Amendola}

\affiliation{INAF/Osservatorio Astronomico di Roma, Via Frascati 33, 00040 Monte
Porzio Catone, Roma, Italy}

\author{Amedeo Balbi}

\affiliation{Dipartimento di Fisica, Universit\`{a} di Roma {}``Tor Vergata'',
via della Ricerca Scientifica 1, 00133 Roma, Italy}

\affiliation{INFN Sezione di Roma {}``Tor Vergata'', via della Ricerca Scientifica
1, 00133 Roma, Italy}

\author{Claudia Quercellini}

\affiliation{Dipartimento di Fisica, Universit\`{a} di Roma {}``Tor Vergata'',
via della Ricerca Scientifica 1, 00133 Roma, Italy}

\date{\today}

\begin{abstract}
It has  been proposed recently to observe the change in cosmological
redshift of distant galaxies or quasars with the next generation of
large telescope and ultra-stable spectrographs (the so-called Sandage-Loeb
test). Here we investigate the possibility of observing the change
in peculiar velocity in nearby clusters and galaxies. This {}``peculiar
acceleration'' could help reconstructing the gravitational potential
without assuming virialization. We show that the expected effect is
of the same order of magnitude of the cosmological velocity shift.
Finally, we discuss how to convert the theoretical predictions into
quantities directly related to observations.
\end{abstract}
\maketitle

\section{Introduction}

One of the most outstanding issues in astrophysics and cosmology is
the measurement of the total mass of clustered structures. The problem
became especially crucial after it was realized that most matter in
the universe is not directly visible through astronomical observations.
Since direct investigation is impossible, a number of indirect techniques
for mass determination have been developed over the years. The most
frequently used methods rely on kinematic measurements, where the
velocity dispersion of some suitable class of test {}``particles''
is used to infer the virial mass of the object under investigation.
This class of methods were successfully applied, for example, to the
rotation curves of spiral galaxies (which led to the first evidence
of the existence of large amounts of dark matter) and to the velocity
dispersion of elliptical galaxies. The largest clustered structures
in the universe where the virial theorem is still applicable are the
clusters of galaxies. The measurement of the velocity of galaxies
in clusters led to further evidence in favour of dark matter, and
so it did the observation of bremsstrahlung x-ray emission from free
electrons of the intracluster gas.

Until now, dynamical methods (i.e. measurements of acceleration) have
not been applied to the measurement of mass, essentially because of
the extraordinary difficulties in detecting tiny velocity variations
for astrophysical objects expected over time scales much smaller than
the cosmic time. However, in the near future we will reach a level
of technological development that might allow the observation of the
dynamical effect of mass. In fact, measuring tiny drifts of the order
of 1 cm/sec in the spectra of astrophysical objects over time scales
of a few decades will be within reach of high-resolution, ultra-stable
spectrographs, coupled to extremely large telescopes (with apertures
of 40-60 m) currently under planning \cite{pasquini06}. Possible
applications of this kind of measurements have been proposed, for
example, to detect extra-solar planets \cite{lovis06}, or for cosmological
applications \cite{sandage62,loeb98}, as a way to detect variations
in the Hubble rate of expansion of the universe, thus investigating
such issues as dark energy or the variation of physical constants
\cite{cora07,balbi07,zhang07}.

In this paper, we explore the possibility of using the time variation
of the peculiar velocity of astrophysical objects (an effect that
we refer to as {\em peculiar acceleration} \cite{loeb98}) as a
way to dynamically determine the mass profile of clustered structures.
This method does not assume virialization, although for simplicity
here we assume a spherically symmetric gravitational potential. We
find numerical estimates of the effect expected for some typical objects,
under idealized conditions, and show that the predicted signal is
of the same order of magnitude of the cosmological one. The determination
of the peculiar acceleration is interesting also because it will act
as a systematic noise source in the observation of the cosmological
velocity shift. Since the detailed specifics for future experiments
are still largely to be finalized, we leave a detailed comparison
of the theoretical predictions with the observational settings to
future work.

\section{Peculiar acceleration}

\begin{figure}[ht!]
\centering \includegraphics[bb=200bp 0bp 450bp 540bp,clip,scale=0.6]{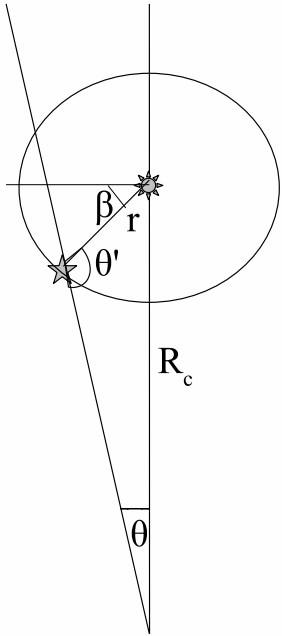}

\caption{Definition of geometric quantities.}

\label{fig:geom} 
\end{figure}

The peculiar acceleration of a particle at the spherical-coordinate
position $(r,\alpha,\gamma)$ in a system centered on a cluster is
\begin{equation}
\vec{a}=\vec{\nabla}\Phi=\Phi_{,r}\hat{r}+\frac{1}{r\sin\gamma}\Phi_{,\alpha}\hat{\alpha}+\frac{1}{r}\Phi_{,\gamma}\hat{\gamma}\label{pecacc}\end{equation}
where the tilded quantities are unit versors and $\Phi$ is the gravitational
potential. We will always assume that Earth's local acceleration has
been properly subtracted. The acceleration along the line of sight
versor $\hat{s}$ is then \begin{equation}
a_{s}=\hat{s}\cdot\vec{\nabla}\Phi,\label{pecacc2}\end{equation}
Let us assume now spherical simmetry for the potential, namely $\Phi=\Phi(r)$.
Therefore the line of sight acceleration is as follows: \begin{equation}
a_{s}=\cos\theta'\Phi_{,r},\end{equation}
where $\theta'$ is as in Fig.~\ref{fig:geom}. In the same figure
we define $R_{c}$ as the cluster distance from observer, $r$ as
the particle distance from cluster's center and $\theta$ the viewing
angle.

For small viewing angles $\theta$ (i.e. for $r\ll R_{c}$) we have
that the peculiar acceleration along the line of sight is \begin{equation}
a_{s}=\cos\theta'\Phi_{,r}\approx\sin\beta\Phi_{,r}|_{r=R_{c}\theta/\cos\beta}\label{accapprox}\end{equation}
and assuming spherical symmetry we obtain \begin{equation}
a_{s}\approx\sin\beta\frac{GM(r)}{r^{2}},\end{equation}
where $r\equiv R_{c}\theta/\cos\beta$. Finally, assuming a density
profile $\rho(r)$, we can write $a_{s}$ in terms of the viewing
angle $\theta$,$R_{c}$, $\beta$ and the density profile parameters.

Let us assume now for definiteness a Navarro-Frenk-White profile (NFW)
\cite{navarro94,navarro95} for the dark matter halo responsible for
the cluster gravitational potential \begin{equation}
\rho(r)=\frac{\delta_{c}\rho_{cr}}{\frac{r}{r_{s}}(1+\frac{r}{r_{s}})^{2}},\label{NFW}\end{equation}
 where $r_{s}=r_{v}/c$ sets the transition scale from $r^{-3}$ to
$r^{-1}$, $c$ is a dimensionless parameter called the concentration
parameter, $\rho_{cr}=3H_{0}^{2}/8\pi G$ is the critical density
at the redshift of the halo ($H_{0}$ being the Hubble constant),
$r_{v}$ is the virial radius inside which the mass density equals
$\Delta_{c}\rho_{cr}$ and $\delta_{c}$ is the characteristic overdensity
for the halo given by \begin{equation}
\delta_{c}=\frac{C\Delta_{c}c^{3}}{3},\label{deltac}\end{equation}
 where\[
C=[\log(1+c)-\frac{c}{1+c}]^{-1}.\]
In addition, $\Delta_{c}$ is the nonlinear density contrast for a
virialized object and enters the expression for the mass $M_{v}=4/3\pi r_{v}^{3}\Delta_{c}\rho_{cr}$.
Its value depends on the cosmological model and assuming a $\Lambda$CDM
we can set $\Delta_{c}=102$ \cite{lahav91}.

Then the mass associated with the radius $r$ is \begin{equation}
M(r)=M_{v}C\Big(\log{(1+\frac{r}{r_{s}})}-\frac{\frac{r}{r_{s}}}{1+\frac{r}{r_{s}}}\Big)\label{mass}\end{equation}
 and consequently \begin{equation}
\Phi(r)_{,r}=\frac{GM_{v}}{r_{s}^{2}}C\Big(\frac{\log{(1+\frac{r}{r_{s}})}}{(\frac{r}{r_{s}})^{2}}-\frac{1}{\frac{r}{r_{s}}(1+\frac{r}{r_{s}})}\Big).\label{pot}\end{equation}
Considering a time interval $\Delta T$ the velocity shift of the
particle test due to the peculiar acceleration along the line of sight
is $s\equiv\Delta v=a_{s}\Delta T$ and turns out to be \begin{equation}
s=\frac{GM_{v}}{r_{s}^{2}}C\Delta T\sin{\beta}\Big(\frac{\log{(1+\frac{r}{r_{s}})}}{(\frac{r}{r_{s}})^{2}}-\frac{1}{\frac{r}{r_{s}}(1+\frac{r}{r_{s}})}\Big).\label{pot}\end{equation}
 Finally we obtain the velocity shift $s(\beta)$ \begin{equation}
s=0.44\frac{\textnormal{cm}}{\textnormal{sec}}\sin\beta\frac{\Delta T}{10yr}\frac{M_{v}}{10^{14}M_{\odot}}(\frac{r_{s}}{1\textnormal{Mpc}})^{2}C\Big(\frac{\log{(1+\frac{r}{r_{s}})}}{(\frac{r}{r_{s}})^{2}}-\frac{1}{\frac{r}{r_{s}}(1+\frac{r}{r_{s}})}\Big)\label{eq:shift}\end{equation}
with $r=R_{c}\theta/\cos\beta$. For a typical cluster value of $C\approx1$,
it turns out therefore that the typical shift for a galaxy cluster
is of the order of 1 cm/sec , similar to the cosmological value at
$z\approx1$ \cite{balbi07} . The maximal value ($\beta\to\pi/2$,
$r\to0$) is : \begin{equation}
s_{0}=0.44\frac{\textnormal{cm}}{\textnormal{sec}}\frac{\Delta T}{10y}\frac{M_{v}}{10^{14}M_{\odot}}(\frac{r_{s}}{1\textnormal{Mpc}})^{2}\frac{C}{2}.\label{eq:shift2}\end{equation}

In the following we estimate the strength of this velocity shift effect
for different clustered objects, i.e. at different distances and scale.
We will always assume $\Delta T=10$yr and let $\beta$ vary in $(0,\pi/2)$.
For $\beta\in(-\pi/2,0)$ the effect is clearly opposite in sign. 

\subsection{Virgo cluster}

\begin{figure}[ht!]
\centering \includegraphics[scale=0.7]{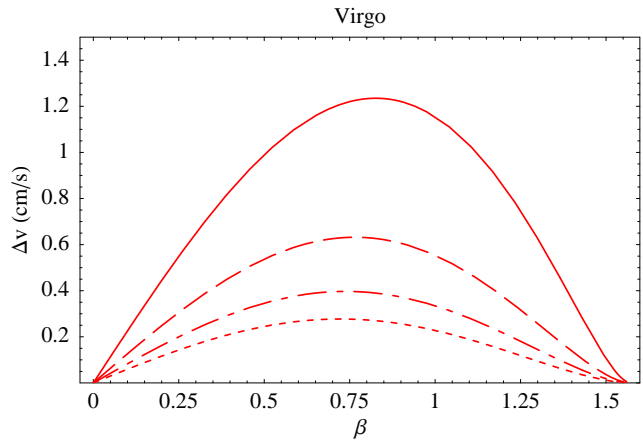}

\caption{{\small Predictions for $\Delta v$ in cm/sec for Virgo ($R_{c}=15$Mpc,
$r_{v}=2.2$Mpc , $c=4$) for four values of $\theta$ up to $\theta_{max}=0.15$
equally spaced, starting from $\theta_{min}=\theta_{max}/4$ (top
to bottom). This value of $\theta_{min}$ corresponds to a radius
of $0.55$Mpc}.}

\label{fig:virgo} 
\end{figure}

The Virgo cluster is the nearest cluster of galaxies. It is also the
most remote cosmic object with a physical connection to our own small
group of galaxies, the Local Group. Its distance is about $15$Mpc
\cite{schmidt07}, corresponding to a redshift of $0.003$ (assuming
that $H_{0}=72$km/s/Mpc): this ensures that the cosmological velocity
shift signal does not come into conflict with the one due to peculiar
acceleration (the cosmological velocity shift being of the order of
$10^{-2}$cm/sec at this redshift). The virial radius is taken to
be $r_{v}=2.2$Mpc corresponding to a virial mass of $1.2\times10^{15}M_{\odot}$;
the concentration parameter is $c=4$, leading to $r_{s}=0.55$ Mpc.
In Fig.~\ref{fig:virgo} the velocity shift as a function of $\beta$
is shown for this cluster. In particular the curve is plotted for
four different values of the viewing angle $\theta$.

\subsection{Coma cluster}

\begin{figure}[ht!]
\centering \includegraphics{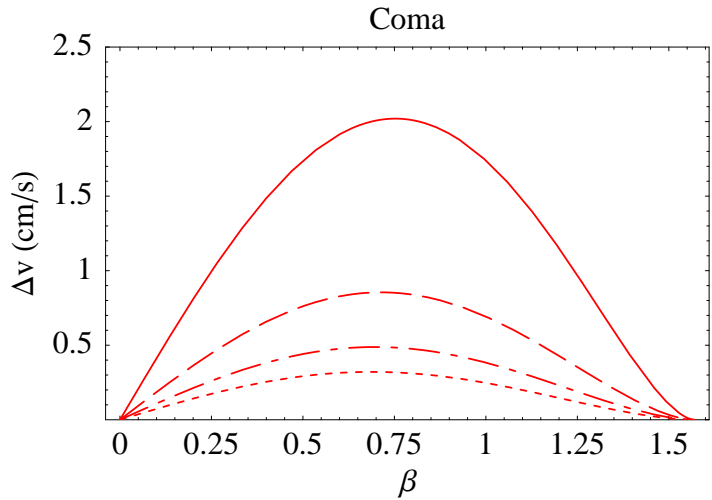}

\caption{{\small Predictions for $\Delta v$ in cm/sec for Coma ($R_{c}=100$Mpc,
$r_{v}=2.7$Mpc , $c=9.4$) for four values of $\theta$ up to $\theta_{max}=0.027$
equally spaced, starting from $\theta_{min}=\theta_{max}/4$ (top
to bottom). This value of $\theta_{min}$ corresponds to a radius
of $0.675$Mpc.}}

\label{fig:coma} 
\end{figure}

Coma is the most studied and best known cluster of galaxies, having
three very appealing properties \cite{schmidt07,lokas03}: it is almost
perfectly spherically symmetric, very rich ($M_{v}=1.2\times10^{15}M_{\odot}$),
and close to our Local Group ($R_{c}=100$Mpc, $z=0.02$ ; the cosmological
velocity shift is of the order of $10^{-1}$cm/sec at this redshift).
Assuming, as above, a NFW density distribution, the other parameters
are: $r_{v}=2.7$ Mpc, $c=9.4$ and $r_{s}=0.29$ Mpc \cite{lokas03}.
The velocity shift predicted for Coma is shown in Fig.~\ref{fig:coma}.

\subsection{Andromeda}

It may be interesting to apply this method on a typical galaxy scale.
We chose to investigate as a representative case the Andromeda galaxy,
located nearly at the center of our Local Group, at a distance of
0.8 Mpc from us. In order to fit the NFW profile to a galaxy, its
parameters must be chosen in a way that does not directly reflect
the immediate physical meaning. For Andromeda, we chose $R_{s}=16$kpc,
$r_{v}=200$ kpc, and $c=12.5$ (\cite{seigar06}). Its mass is estimated
to be roughly $M=10^{12}M_{\odot}$. Obviously in this case the test
particles will be represented by stars or gas clouds, rather than
galaxies.

\begin{figure}[ht!]
\centering \includegraphics{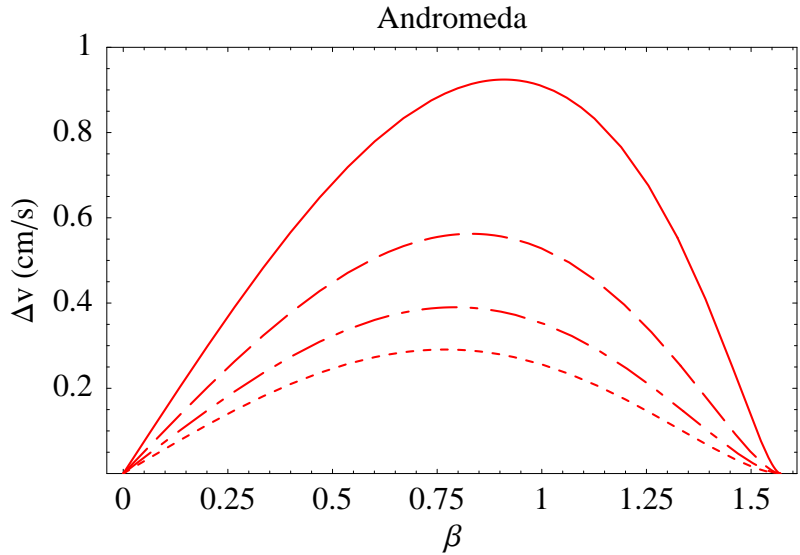}

\caption{{\small Predictions for $\Delta v$ in cm/sec for Andromeda ($R_{c}=0.8$Mpc,
$r_{v}=200$kpc , $c=12.5$) for four values of $\theta$ up to $\theta_{max}=0.25$
equally spaced, starting from $\theta_{min}=\theta_{max}/4$ (top
to bottom). This value of $\theta_{min}$ corresponds to a radius
of $7.5$kpc. The signal increases for smaller radius.}}

\label{fig:andro} 
\end{figure}

\section{Observing the peculiar acceleration}

Since observations are performed along the line of sight, it is practically
impossible to obtain information on the galaxy position within the
cluster, i.e. on the $\beta$ angle. It is then useful to identify
some observable quantities that can be directly compared to theoretical
predictions. We propose two such quantities in the following subsections.

\subsection{Maximum velocity shift}

\begin{figure}[ht!]
\centering \includegraphics[scale=0.7]{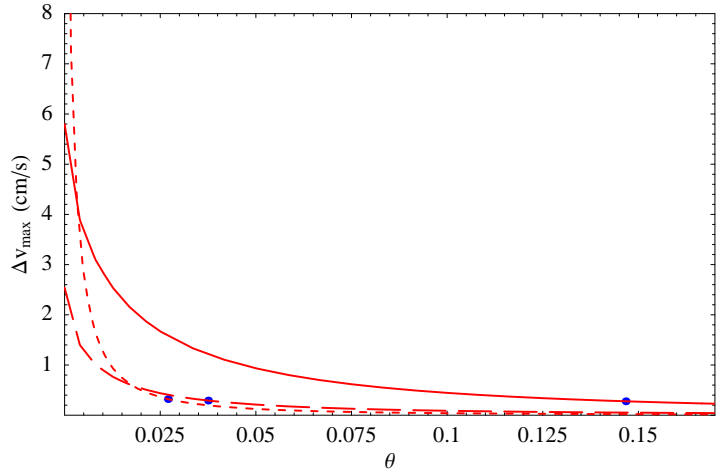}

\caption{{\small The maximum of the velocity shift due to peculiar acceleration
as a function of the viewing angle $\theta$ for Virgo (straight line),
Coma (short-dashed line) and Andromeda (long-dashed line). The round
points mark the value of $\theta_{max}$ lying on the correspondent
curves.}}

\label{fig:tot_max} 
\end{figure}

\begin{figure}[ht!]
\centering \includegraphics[scale=0.7]{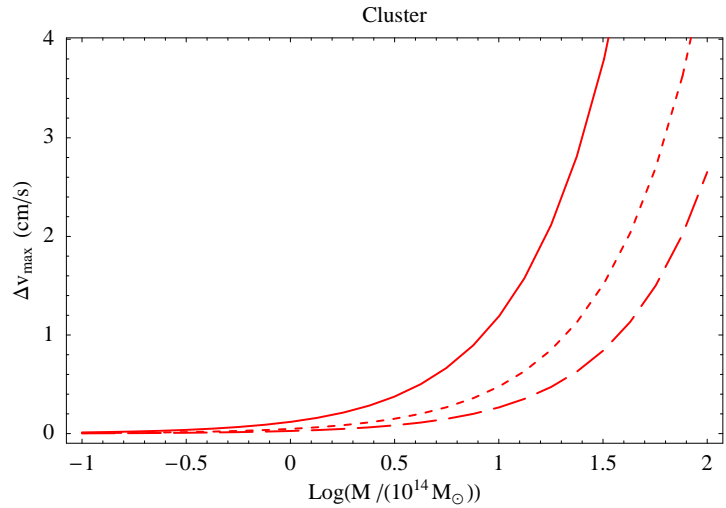}

\caption{{\small The velocity shift signal as a function of mass for a Coma-like
cluster, located at the same distance as Coma, for three different
values of $\theta$ separated by $\theta_{max}/3$, starting from
$\theta_{max}/3$ (top to bottom). }}

\label{fig:max_c} 
\end{figure}

\begin{figure}[ht!]
\centering \includegraphics[scale=0.7]{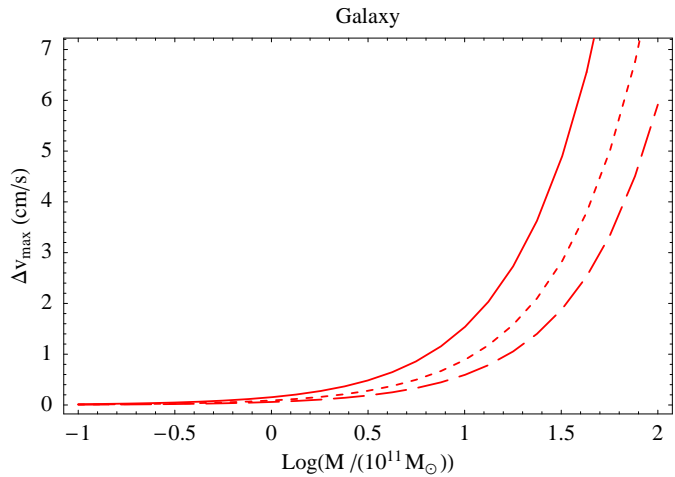}

\caption{{\small The velocity shift signal as a function of mass for an Andromeda-like
galaxy, located at the same distance as Andromeda, for three different
values of $\theta$ separated by $\theta_{max}/3$, starting from
$\theta_{max}/3$ (top to bottom). }}

\label{fig:max_g} 
\end{figure}

The predicted velocity shift reaches a maximum value along the line
sight, $\Delta v_{max}$, that depends on the halo profile. By estimating
$\Delta v_{max}$ from a sample of galaxies along each line of sight,
one can therefore constrain the cluster gravitational potential. Fig.~\ref{fig:tot_max}
shows the behaviour of $\Delta v_{max}$ as a function of $\theta$
for Virgo, Coma and Andromeda. We note that each curve can be actually
observed only for angles up to $\theta_{max}$. We also plot in Fig.~\ref{fig:max_c}
and Fig.~\ref{fig:max_g} the same quantity $\Delta v_{max}$ as
a function of mass for a cluster and a galaxy, respectively. Clearly
the signal is independent of the object distance for a maximal viewing
angle $\theta_{max}\sim1/R_{c}$.

\begin{figure}[ht!]
\centering \includegraphics{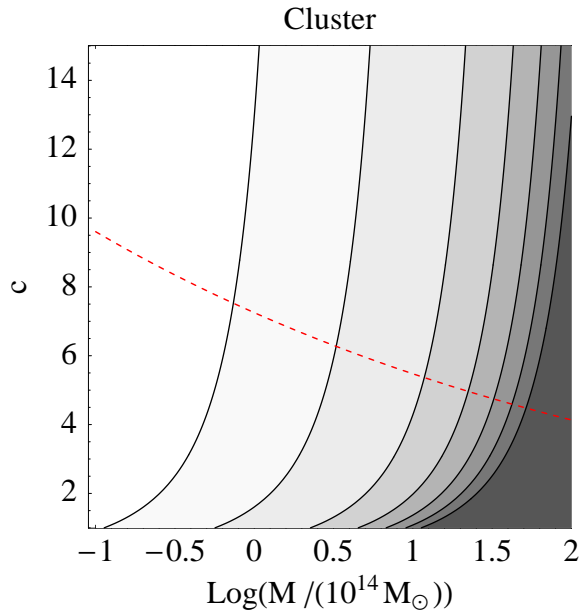}

\caption{{\small Contour plot for the velocity shift of a cluster at $\theta_{max}/4$
as a function of the concentration $c$ and the mass $M$. Higher
values correspond to darker regions and the contours levels are (left
to right) $\Delta v=(0.1,0.5,2,4,6,8,10,15)$cm/s. The short-dashed
line is the fitting formula by \cite{bullock01}. }}

\label{fig:contour_cluster} 
\end{figure}

\begin{figure}[ht!]
\centering \includegraphics{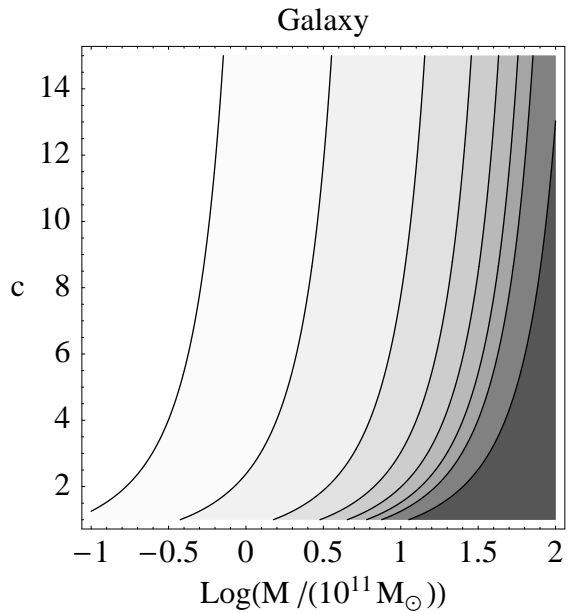}

\caption{{\small Contour plot for the velocity shift of a galaxy at $\theta_{max}/4$
as a function of the concentration $c$ and the mass $M$. Higher
values correspond to darker regions and the contours levels are (left
to right) $\Delta v=(0.1,0.5,2,4,6,8,10,15)$cm/s.}}

\label{fig:contour_galaxy} 
\end{figure}

In Fig.~\ref{fig:contour_cluster} and Fig.~\ref{fig:contour_galaxy}
we plot contour curves corresponding to the same $\Delta v_{max}$
for $\theta=\theta_{max}/4$ and different combinations of the concentration
parameter $c$ and mass. In principle, knowing the mass of the object
from other observations, a measurement of the velocity shift can give
indications on the concentration parameter. In particular, recently
several authors have derived fitting formulae for the concentration
parameter as a function of mass from $N$-body simulations (\cite{bullock01,eke01,maccio07}).
We show a curve from one of these formulae in Fig.~\ref{fig:contour_cluster}.
Then, for clusters, a measurement of $\Delta v_{max}$ at different
mass scales would provide a test of the concentration parameter fitting
formula. Clearly, instead of the maximum one could use the variance
of the peculiar acceleration or another non-vanishing average. In
the next subsection we use instead the full distribution to relate
the theoretical prediction to observable quantities.

\begin{figure}[ht!]
\centering \includegraphics{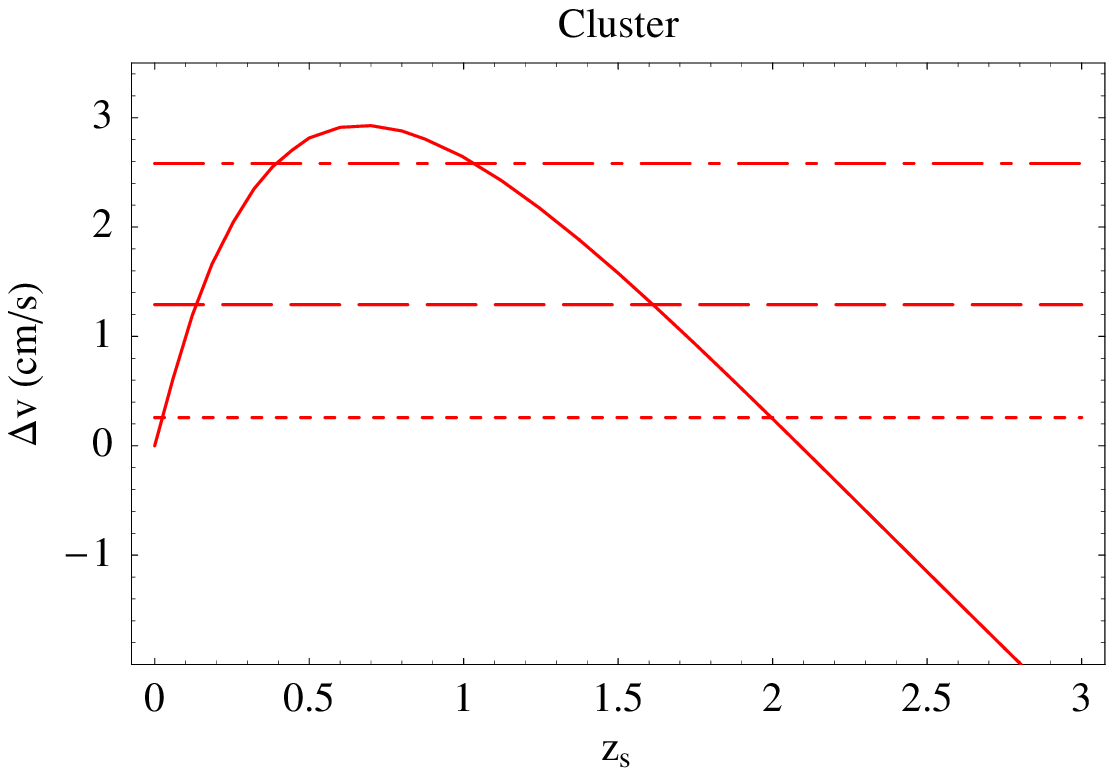}

\caption{The cosmological velocity shift as a function of redshift for a $\Lambda$CDM
model (solid line) and the maximum of the velocity shift due to the
peculiar acceleration for $\theta=\theta_{max}/4$ and three different
value of the mass of a rich cluster: $M=10^{14}M_\odot$ (short-dashed
line), $M=5\cdot10^{14}M_\odot$ (long-dashed line), $M=10^{15}M_\odot$
(long-short-dashed line).}

\label{fig:cosmo_clus} 
\end{figure}

\subsection{Density distribution of galaxies}

Knowing the radial distribution of galaxies $\rho_{g}(r)$ (i.e. deriving
it from the projected number density) one can compare the observed
distribution $N(\Delta s)$ (number of galaxies in the velocity shift
bin $\Delta s$) with the prediction. 

The expected number of galaxies in an annulus $\Delta\theta$ and
thickness $\Delta(R_{c}\theta\tan\beta)=R_{c}\theta(1/\cos^{2}\beta)\Delta\beta$
around the cluster center is\begin{equation}
n(R_{c},\theta,s)\Delta\theta\Delta s=2\pi\rho_{g}\left(\frac{R_{c}\theta}{\cos\beta}\right)\frac{R_{c}^{3}\theta^{2}}{\cos^{2}\beta}\Delta\theta\Delta\beta\end{equation}
 and therefore we expect in the interval $\Delta s$ a number of galaxies
proportional to\begin{equation}
n(R_{c},\theta,s)\Delta\theta\Delta s=2\pi\rho_{g}\left(\frac{R_{c}\theta}{\cos\beta}\right)\frac{R_{c}^{3}\theta^{2}}{\cos^{2}\beta}\Delta\theta\frac{\Delta s}{(ds/d\beta)}\label{eq:obs}\end{equation}
 where $\beta=\beta(s)$ can be obtained by inverting (\ref{eq:shift}).
The histogram $N(R_{c},\theta,s)=n\Delta s$ is therefore given by\begin{equation}
N(R_{c},\theta,s)=2\pi\rho_{g}\left(\frac{R_{c}\theta}{\cos\beta(s)}\right)\frac{R_{c}^{3}\theta^{2}}{\cos^{2}\beta(s)}\frac{d\beta}{ds}\Delta s\label{eq:obs3}\end{equation}

\begin{figure}[ht!]
 \centering \includegraphics{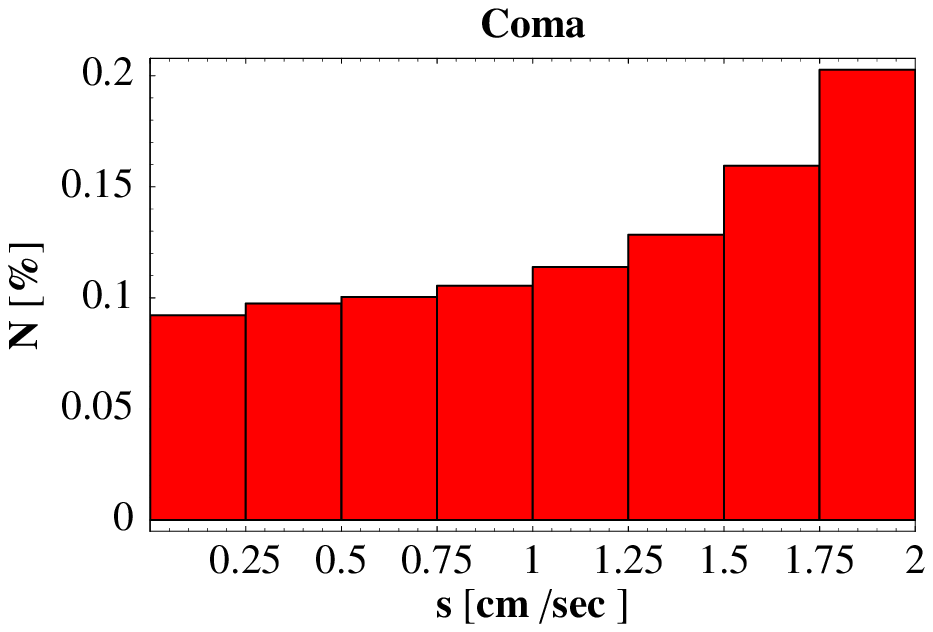}

\caption{Frequency histogram $N(s)$ (expected frequency of galaxies in velocity
shift bins) for Coma at $\theta=\theta_{max}/4$ . Here we assumed
that the radial distribution of galaxies is proportional to the halo
profile. }

\label{fig:distr} 
\end{figure}

Now Eq. (\ref{eq:obs3}) contains only observable quantities. In principle,
confronting this expected number $N$ with the observed numbers one
can reconstruct $\rho(r)$. An example of $N(s)$ is given in Fig.
(\ref{fig:distr}).

\subsection{Comparing with the cosmological signal }

As already mentioned in the previous section, the cosmological signal
is competitive with the velocity shift due to the peculiar acceleration.
In Fig. \ref{fig:cosmo_clus} the cosmological signal for a $\Lambda$CDM
model with $\Omega_\Lambda=0.7$ is plotted together with the maximum
of the velocity shift of a cluster; the redshift of their \char`\"{}equivalence\char`\"{}
is very small ($z\simeq 0.02$) depending on the mass of the cluster.
In addition, there is a second redshift, around $z\simeq 2$, at which
the cosmological signal is negligible. However observations at such
a distance could not be reliable, not only because of the difficulty
of precise measurements at this redshift, but also because the zero-crossing
of the function is cosmology-dependent. In principle of course the
cosmological signal could also be subtracted from the peculiar one
after averaging over a consistent number of galaxies.

\section{Conclusions}

In this paper, we performed a first exploration of the magnitude of
peculiar acceleration expected in clustered structures in the universe,
over time scales of one or a few decades. Our study was motivated
by the interesting prospects of observing velocity shifts of cosmic
objects of order of a few cm/sec in a decade offered by forthcoming
extremely large telescopes coupled to high-resolution ultra-stable
spectrographs. Peculiar acceleration would be an interesting target
for such future measurements, since, as we have shown, could give
an independent measurement of the mass profile of clusters of galaxies,
and possibly even of individual galaxies. We have also shown that
measurements of the velocity shift for different mass scales could
test the relation between the concentration parameter of clusters
and the virial mass.

Our results show that the expected signals for the objects we considered
(the Coma and Virgo clusters, and the Andromeda galaxy) are comparable
to the cosmological change in redshift, due to the expansion dynamics,
expected for distant QSO's \cite{pasquini06}. Due to the high stability
of Ly-$\alpha$ lines in QSO's, it has been recently argued that the
cosmological effect could be observed in a time scale of about three
decades \cite{cora07,balbi07}, giving important information on the
composition of the universe, and in particular on the mysterious dark
energy which drives its current accelerated expansion. It seems plausible
that an effect of similar order of magnitude, predicted by our study,
could then be also observable, although the objects used as test particles
are different in nature. Our study is also useful to the extent of
assessing the peculiar acceleration as a source of errors in the measurement
of the cosmological velocity shift \cite{loeb98}.

Of course, we realize that the experimental and observational details
of future instruments are still too blurry to make any clear statement
on the actual feasibility of using peculiar acceleration as an astrophysical
and cosmological tool. Many issues are still to be discussed before
forecasting the results of actual observations, from the stability
of the galactic spectra to the effect of local deviations from spherical
symmetry. However, we believe it is exciting to entertain the possibility
that peculiar acceleration might actually be observable in the not
too distant future, and hope that more detailed investigations of
its applicability in astrophysics might be stimulated by our study.

\bibliographystyle{apsrev} \bibliographystyle{apsrev}
\bibliography{Bacc}

\end{document}